\documentstyle[prb,aps]{revtex}

\begin{document}
\twocolumn[\hsize\textwidth\columnwidth\hsize\csname
@twocolumnfalse\endcsname

\title{Irreversible magnetization under rotating fields and lock-in effect on ErBa$_2$Cu$_3$O$_{7-\delta}$ single crystal with columnar defects}

\author{M. A. Avila$^a$, L. Civale$^b$, A. V. Silhanek$^b$, R. A. Ribeiro$^a$, O. F. de Lima$^a$, H. Lanza$^c$}

\address{$^a$Instituto de F\'{i}sica ''Gleb Wataghin'', UNICAMP, 13083-970, Campinas - SP, Brazil.\\
$^b$Comisi\'on Nacional de Energ\'{\i}a At\'omica - Centro At\'omico Bariloche and Instituto Balseiro, 8400 Bariloche, Argentina.\\
$^c$Comisi\'on Nacional de Energ\'{\i}a At\'omica - Departamento de F\'{\i}sica, Buenos Aires, Argentina.}

\date{June 15, 2001}
\maketitle

\begin{abstract}
We have measured the irreversible magnetization (${\bf M_i}$) of
an ErBa$_{2}$Cu$_{3}$O$_{7-\delta}$ single crystal with columnar
defects (CD), using a technique based on sample
rotation under a fixed magnetic field $H$. This method is valid for samples whose magnetization vector remains perpendicular to
the sample surface over a wide angle range - which is the case
for platelets and thin films - and presents several advantages
over measurements of ${\bf M_L}(H)$ loops at fixed angles. The resulting
${\bf M_i}(\Theta)$ curves for several temperatures show a peak
in the CD direction at high fields. At lower fields, a very well defined
plateau indicative of the vortex lock-in to the CD develops. The H dependence of the lock-in angle $\varphi_{L}$ follows the $H^{-1}$ theoretical prediction, while the temperature dependence is in agreement with entropic smearing effects corresponding to short range vortex-defects interactions.

\end{abstract}

\vskip1pc] \narrowtext

\section{Introduction}

The study of the angular dependence of vortex pinning in high
temperature superconductors (HTSC) with tilted columnar defects
has revealed a richer variety of phenomena and pinning regimes
than originally expected. At high temperatures and magnetic
fields, the uniaxial nature of pinning by CD dominates the vortex
response\cite{civale91}. This is clearly seen, for instance, when
isothermal magnetization loops ${\bf M(H)}$ are measured for
different field orientations\cite{silhanek99a}. At fixed field
modulus $H$, the irreversible magnetization $M_i=\frac 12 \Delta
M$ (where $\Delta M$ is the width of the hysteresis, proportional
to the persistent current density $J$) exhibits a well defined
maximum when ${\bf H} \parallel$ CD. For other orientations
``staircase vortices'' develop. In a previous study we have
shown\cite{silhanek99a} that in YBa$_2$Cu$_3$O$_7$ (YBCO), and
due to the simultaneous presence of CD, twin boundaries and
crystallographic ab-planes, correlated pinning dominates over
random pinning for all orientations, forming staircases of
different configuration (i.e., with segments locked into
different correlated structures) depending on the field direction.

An additional feature is the existence of a lock-in phase. When
the angle between ${\bf H}$ and the CD is less than a lock-in
angle $\varphi_{L}\left(H,T\right)$, it is energetically
convenient for vortices to ignore the ${\bf H}$ orientation and
to remain locked into the tracks\cite{nel-vin,blatter94}. Since $\varphi_{L}$ scales as
$1/H$, in practice this effect is only visible at low fields. An
experimental manifestation of the lock-in regime is the
existence\cite{silhanek99a} of a ``plateau'' in the irreversible
magnetization, $M_i\left(\Theta\right) \approx const$, over a
certain angular range. Here $\Theta$ is the angle between the normal to the platelet crystal, ${\bf n}$ (which coincides with the crystallographic c-axis) and ${\bf H}$, defined within the
plane that contains the CD.

At low fields, an additional effect must be taken into account.
Due to both the anisotropic superconducting response of the HTSC
and the sample geometry, the direction of the internal field
${\bf B}$, that coincides with the direction of the vortices,
differs from that of ${\bf H}$. As the uniaxial pinning of the CD
maximizes when ${\bf B}$ (rather than ${\bf H}$) is aligned with
the tracks, the maximum in $M_i\left(\Theta\right)$ occurs at an
angle that progressively departs\cite{silhanek99a} from the
orientation of the tracks, $\Theta_D$, as H decreases.

The low field misorientation between ${\bf B}$ and ${\bf H}$ poses
a serious experimental concern. All studies of the pinning
properties of tilted CD that are based solely on measurements at
${\bf H} \parallel$ CD, or on comparison of this orientation with
a few others, give valid information at high fields, but are
misleading at low fields; vortices are just not oriented in the
right direction. To avoid this problem, a rather complete
knowledge of the angular dependent response, $M_i \left(H, \Theta
\right)$, is required.

In this work we present a procedure that allows us to obtain
directly $M_i\left(\Theta\right)$ by rotating the sample at fixed
$H$ and $T$. This method has the advantage that a fine grid can be easily
obtained in the angular ranges of interest, thus permitting the
exploration of the various regimes with significantly improved
angular resolution. We apply this experimental procedure to
investigate the pinning produced by tilted CD in an
ErBa$_2$Cu$_3$O$_7$ single crystal. We present a detailed
analysis of the lock-in angle as a function of $H$ and $T$. The
width of the lock-in regime is shown to follow a $1/H$ dependence
over a wide temperature range, and from the temperature dependence
of the slope of $\varphi_{L}$ vs $1/H$ we determine the entropic smearing function $f\left(T/T^*\right)$.

\section{Experimental Details}

The sample used in this work is a rectangular
ErBa$_2$Cu$_3$O$_{7-\delta}$ single crystal platelet of
dimensions $0.44\times 0.33\times 0.01 mm^3$, grown by the self-flux method in a commercial
yttria-stabilized-zirconia crucible \cite{avila00}. After growth
it was annealed under oxygen atmosphere for 7 days at
450$^{\circ}$C. This sample was then irradiated with 309 MeV
$Au^{26+}$ ions (whose penetration range in this material is $\sim 15\mu m$) at the TANDAR accelerator in Buenos Aires (Argentina), to introduce columnar defects at an angle
$\Theta_D\approx 30^{\circ}$ off the {\it c} axis. The rotation
axis, which is perpendicular to the plane formed by the c-axis
and the track's direction, is parallel to the largest crystal
dimension. The irradiation dose was equivalent to a matching
field of $B_{\Phi} = 1 T$. After irradiation the sample presented
a superconducting transition temperature of $T_c = 90.0 K$ and
transition width of $\Delta T < 1 K$.

Magnetization experiments were conducted on a commercial
superconducting quantum interference device (SQUID) magnetometer
(Quantum Design MPMS-5), equipped with two sets of detectors that
allow to record both the longitudinal ($M_L$) and transverse
($M_T$) components of the magnetization vector ${\bf M}$, with
respect to the longitudinally applied field ${\bf H}$. We have
developed a sample rotation system (hardware and software) that
solves the problems usually involved in the measurement of $M_T$,
and thus allows us to study the response of the samples at
arbitrary orientations. We have used that system in the past to
measure magnetization loops $\bf {M}(H)$ at different field
orientations in samples similar to the one investigated here.\cite{silhanek99a,anomalous} In
those experiments the sample is initially rotated to the desired
$\Theta$, then zero-field-cooled (ZFC) from above $T_c$ to the
desired measuring temperature $T$. Both $M_L(H)$ and $M_T(H)$ are
then measured, and the separation between the upper and lower
branches of both loops ($\Delta M_L(H)$ and $\Delta M_T(H)$) is
used to calculate the amplitude $M_i=\frac 12 \sqrt{\Delta
M_L(H)^2 + \Delta M_T(H)^2}$ and direction
$\Theta_M=arctan(\Delta M_T(H)/ \Delta M_L(H))$ of the
irreversible magnetization vector ${\bf M_i}$.

The alternative rotating sample measurements presented here are
performed by setting up a desired initial state $\left(
T,H,\Theta_H \right)$ and then recording $M_L(\Theta)$ and
$M_T(\Theta)$ for fixed $T$ and $H$. The sample is rotated a
given angle step (typically 1$^{\circ}$ to 3$^{\circ}$) and
re-measured. Usually, the procedure is repeated until the crystal
completes 2 or 3 full turns. This provides us with redundant
information that contributes to improve the quality of the data.
After careful subtraction of the signal of the plastic sample
holder (which has only longitudinal component and is small,
linear in $H$, almost temperature independent and, most
importantly, angle independent) and of the reversible response,
the irreversible components $M_{Li}(H)$ and $M_{Ti}(H)$ are used
to determine $M_i= \sqrt{M_{Li}(H)^2 + M_{Ti}(H)^2}$ and
$\Theta_M=arctan(M_{Ti}(H)/M_{Li}(H))$.

\section{Rotating measurements}

\subsection{Meissner response}

We began this study with an analysis of the Meissner response. To
that end we ZFC the crystal, then applied a field $H$ smaller than
the lower critical field $H_{c1}\left(\Theta\right)$ for all
$\Theta$, and subsequently performed the rotating measurements.
Ideally, under those conditions there are no vortices in the
crystal and the response depends neither on the material
anisotropy nor on the pinning properties, it is totally
determined by the sample geometry.  As was previously
shown\cite{candia99}, $M_L\left(\Theta\right)$ and
$M_T\left(\Theta\right)$ for a thin platelet should follow the
dependencies

\begin{equation}
4\pi M_L\left(\Theta\right)=-H\left( \frac 1{2\nu }\cos ^2 \Theta+\frac 1{1-\nu }\sin^2 \Theta\right) \label{eq:meissL1}
\end{equation}

\begin{equation}
4\pi M_T\left(\Theta\right)=-H\left( \frac 1{2\nu }-\frac 1{1-\nu }\right) \sin \Theta\cos \Theta \label{eq:meissT1}
\end{equation}
where $\nu$ is the appropriate demagnetizing factor, which is essentially given by the thickness of the platelet ($t$) divided by its width ($W$). These equations can be easily rewritten as

\begin{equation}
M_L=-M_0-M_{2\Theta }\cos 2\Theta_H \label{eq:meissL2}
\end{equation}

\begin{equation}
M_T=-M_{2\Theta }\sin 2\Theta_H \label{eq:meissT2}
\end{equation}
where

\begin{equation}
4\pi M_0=\frac H2\left( \frac 1{2\nu }+\frac 1{1-\nu }\right) \label{eq:meissM0}
\end{equation}

\begin{equation}
4\pi M_{2\Theta}=\frac H2\left( \frac 1{2\nu }-\frac 1{1-\nu }\right) \label{eq:meissM2}
\end{equation}

Equations \ref{eq:meissL2} and \ref{eq:meissT2} indicate that the
magnetization vector {\bf M} can be visualized as the sum of a
fixed contribution ${\bf M_0}$, anti-parallel to ${\bf H}$, and a
rotating contribution ${\bf M_{2\Theta}}$ with a periodicity of
$180^{\circ}$. This suggests that a convenient way to plot these
data is on an $M_L$,$M_T$ plane. In this presentation the Meissner
response is expected to lie on a circumference of radius
$M_{2\Theta}$ centered at
$\left(M_L,M_T\right)=\left(-M_0,0\right)$. One complete
circumference is drawn by a rotation of $180^{\circ}$. An example
of this procedure (for $T=60K$ and $H=50Oe$) is shown in Figure
1. The crystal was rotated by two complete turns, thus there are
four sets of data points covering $180^{\circ}$ each, which are
clearly separated in two groups. This is due to a small remnant
magnetization ${\bf M_R}$, which originates from the small
residual field that is usually present during the ZFC.\cite{candia99}

The vector ${\bf M_R}$ has fixed modulus and its direction
remains fixed with respect to the sample during rotation,\cite{candia99} thus
$M_{RL}=M_R\cos \left( \Theta+\Theta_R \right)$ and $M_{RT}=M_R\sin
\left( \Theta+\Theta_R \right)$, where $M_R$ and $\Theta_R$ are
constants. Since ${\bf M_R}\left(\Theta\right)$ has a one fold
periodicity, it breaks the Meissner two fold periodicity and
splits the experimental data into two sets. Indeed, by fitting
the data in Fig. 1 using a combination of the Meissner and
remnant contributions we can easily determine and remove the
remnant part and all points collapse on a single circumference
(solid symbols). Fig. 1 is an extreme example of remnant
influence, chosen to show that even in that case the Meissner
response can be obtained. By carefully canceling the residual
magnetic field in the ZFC procedure we can obtain a much smaller
${\bf M_R}$ such that both circumferences of raw data in Fig. 1
almost collapse on a single one.

Equations \ref{eq:meissL2} and \ref{eq:meissT2} were used to fit
several measurements for different temperatures and fields, and
used to calculate the sample volume $V\approx 1.45\times 10^{-6}
cm^3$ and demagnetization factor $\nu\approx 0.033$. Both results
are in very good agreement with the values directly determined from crystal dimensions (V $\sim 1.46 \times 10^{-3} cm^3; \nu \approx 0.03$).

\subsection{critical state}

We now focus on the high field range, where the crystal is in the
mixed state. Fig. 2(a) shows $M_L(\Theta)$ and $M_T(\Theta)$ for a
rotation at $70K$ and $8kOe$, where the angle independent
background due to the holder has already been removed from $M_L$.
As the reversible magnetization of the superconductor [$\sim
\left( \Phi_0 /32\pi^2\lambda^2 \right) \ln \left( H_{c2}/H
\right) \sim 5G$] is negligible compared to $\bf{M_i}$, the
response is dominated by vortex pinning. Curves in Fig. 2(a)
exhibit a rich structure, due to the combination of crystalline
anisotropy, directional vortex pinning and geometrical effects.
In order to extract useful information from them, we must first
establish the relation between $\bf{M_i}$ and the screening
current $\bf J$ flowing through the crystal.

For simplicity, we will analyze the case of a thin infinite strip
of aspect ratio $\nu=t/W \ll 1$, that can rotate around its axis,
which is perpendicular to $\bf H$. Let's assume that the strip
was originally ZFC at an angle $\Theta$ and $H$ was subsequently
applied (the initial condition in Fig. 2(a)). If $H$ is high
enough we can consider\cite{clem-sanch} that a current
density of uniform modulus $J_c(\Theta)$ flows over the whole
volume.\cite{clem-sanch} This $\bf J$ is parallel to the strip axis and it reverses
sign at the plane that contains the axis and $\bf H$.

It has been shown\cite{prozorov96,zhukov97,hasanain99} that, in
this fully penetrated critical state and as long as
$\nu\tan\Theta \ll 1$, the angle between ${\bf M_i}$ and the
sample normal $\bf n$ is $\alpha \sim \arctan \left(\frac
{2}{3}\nu^2\tan\Theta\right) \ll \Theta$. That is, ${\bf M_i}$
remains almost locked to $\bf n$ due to a purely geometrical
effect. For the particular crystal of the present study, $\alpha$
should be smaller than $1^{\circ}$ for $\Theta \le 80^{\circ}$.
Another result\cite{zhukov97} is that, although in principle the
geometrical factor relating $M_i$ with $J_c(\Theta)$ depends on
$\Theta$, within that same angular range the variations are given
by the factor $\left(1-\frac {2}{3}\nu^2\tan^2\Theta\right)$ and
thus are negligible.

We now discuss what happens when the strip is rotated away from
this initial state by a small angle $\delta \Theta$. The result
will depend on the direction of rotation. If $\bf n$ approaches
$\bf H$ (this corresponds to the angular ranges $90^{\circ}$ to
$180^{\circ}$ and $270^{\circ}$ to $360^{\circ}$ in Fig. 2(a)),
the normal component $H_{\perp}$ will increase, thus inducing
screening currents at the edges of the crystal in the same
direction as those already flowing. Vortices will then displace
to satisfy the condition $J \le
J_c\left(\Theta+\delta\Theta\right)$ everywhere. If
$J_c\left(\Theta+\delta\Theta\right)\le J_c(\Theta)$ the new
distribution will be analogous to the initial one, with
$J=J_c(\Theta+\delta\Theta)$ everywhere and the boundary of
current reversal rotated by an angle $\delta \Theta$ in order to
remain parallel to $\bf H$. On the contrary, if
$J_c(\Theta+\delta\Theta) > J_c(\Theta)$, the new field profile
will propagate all the way to the center of the sample only if
$\delta H_{\perp} = H \sin(\Theta)\delta\Theta$, is larger than
the maximum possible additional screening $\sim t\left[
J_c(\Theta+\delta\Theta)-J_c(\Theta) \right]$. The condition for
the ``full penetration of the rotational perturbation'' is thus

\begin{equation}
H \sin(\Theta) \ge t \frac {dJ_c}{d\Theta} \label{eq:fullp}
\end{equation}

If the inequality (\ref{eq:fullp}) is satisfied, the vortex
system will evolve under rotations maintaining a fully penetrated
critical state with uniform $J$. In other words, the state at any
$\Theta$ will be the same that would have formed by increasing
$H$ after ZFC at that orientation. Then, as long as $\nu\tan\Theta
\ll 1$, the condition that ${\bf M_i}$ is almost parallel to $\bf
n$ is preserved. We have experimentally confirmed this fact:
$\alpha \le 1^{\circ}$ for all measurements conducted in this
work, except in a very narrow angular range around the ab-planes,
where a flip in ${\bf M_i}$ occurs.\cite{zhukov97} Thus, from now on we will
plot all the results as a function of $\Theta$. Furthermore, we
can obtain $J_c(\Theta)$ by simply multiplying $M_i$ by the angle
independent factor that corresponds to the relation valid for
$\bf H \parallel \bf n$. If eq. (\ref{eq:fullp}) is not
satisfied, $J$ will be subcritical in part of the sample and this
relation is no longer valid.

If the crystal is rotated in such a way that $\bf n$ moves away
from $\bf H$ (so $H_{\perp}$ decreases), the new screening
currents induced at the edges of the crystal will oppose to those
already flowing. As the rotation progresses the boundary between
the old and new $\bf J$ directions will move inwards, until
eventually the new critical state propagates to the whole sample.
From that point the situation will again be analogous to that
already discussed, except that ${\bf M_i}$ will be paramagnetic
instead of diamagnetic.

A rotation at fixed $H$ is to some extent analogous to a
hysteresis loop\cite{prozorov96}. Rotating $\bf n$ towards $\bf H$
increases $H_{\perp}$, which is roughly equivalent to increasing
$H$ at $\Theta=0^{\circ}$, moving along the lower (diamagnetic)
branch of the loop. Decreasing $H_{\perp}$ (either by rotating
$\bf n$ away from $\bf H$ or by crossing the $\bf H \parallel \bf
c$ condition), is equivalent to reversing the field sweep, thus
producing a switch to the other branch of the loop. This is a
useful analogy for the analysis of the rotations, although it
should not be pushed too far.

A basic difference is that a rotation also produces a variation in
the parallel field component, $\delta
H_{\parallel}=H\cos(\Theta)\delta\Theta$. This generates
screening currents flowing in opposite directions on the upper and
lower surfaces of the strip, which produce a tilting force on the
vortices.\cite{clem82,clem86,perez90} If the perturbation propagates all the way to the
central plane, the result is a rotation of the vortex direction
following $\bf H$, the situation that we have implicitly assumed
above. However, if pinning were strong enough it could preclude
the propagation of the tilt beyond a certain depth, thus
generating a critical state along the crystal thickness, with a
central segment of the vortices remaining in the original
direction.\cite{clem82,clem86,perez90,goeckner94,hasanain96,hasan97,obaidat97} If this
effect were significant, as the rotation proceeded the
orientation of the vortices would lag behind the field direction.
In an extreme case, vortices deep inside the sample would rotate
rigidly with it, a situation that has indeed been
observed\cite{vlasko98,obaidat98,hasan99}. As we will show below,
in the present case we have clear experimental evidence that the
misorientation between the vortex direction and $\bf H$ due to
this lag effect is negligible, so all this complication can be
ignored.

We now analyze the curves shown in Fig. 2. The measurement starts
at $\Theta \sim 30^{\circ}$ (point A) with $\bf n$ rotating away
from $\bf H$. Thus, $\bf J$ initially undergoes a flip until the
reversed fully penetrated critical state is formed (point B).
From here the evolution of the system turns independent of the
initial conditions and becomes two fold periodic. From point C
($\Theta = 90^{\circ}$) to point E ($\Theta = 180^{\circ}$) the
system evolves in a fully penetrated critical state (in the
hysteresis loop analogy, this is equivalent to increasing the
field from zero to $H$). Clearly visible within this angular
range is the peak in both $M_L$ and $M_T$ at $\Theta \sim
150^{\circ}$ (point D), that corresponds to the direction of the
CD. At point E, $M_T$ is null as expected by symmetry, while $M_L$
begins a quick flip due to the reversal of the screening currents
as $H_{\perp}$ reaches a maximum at $\bf H
\parallel \bf n$ and then starts to decrease. The end of this
flip at point G indicates that the critical state is completely
reversed. From G to C' ($\Theta=270^{\circ}$) the evolution is
analogous to a field decreasing portion of a loop.

Note that between E and G there is one unique angle (point F)
where both $M_L$ and $M_T$ are null. This condition is equivalent
to the unique $H$ value in the switch from the lower to the upper
branch of a $M(H)$ loop where ${\bf M_i=0}$. The fact that the
condition $M_L=0$ occurs at the same angle where $M_T=0$ confirms
that the background signal has been correctly subtracted, and we
have systematically made use of this checking procedure.

In Fig. 2(a) the direction of rotation is such that the conditions
$\bf H \perp c$; $\bf H \parallel$ CD and $\bf H \parallel$ c proceed in that order.
We define this as a clockwise (CW) rotation. In contrast, in
a counter-clockwise (CCW) rotation the alignment occurs when $\bf
n$ is moving away from $\bf H$. The consequences of this
difference are described below.

In Figure 2(b) the same CW data of Fig. 2(a) is shown in an $M_L$
vs $M_T$ polar graph (full symbols), together with the CCW
rotation under the same conditions (open symbols). In both cases
the initial behavior until the critical state is fully developed
(portion A to B in the CW and P to Q in the CCW) and the
subsequent $180^{\circ}$-periodic evolution in the critical state
(covering approximately two periods of $180^{\circ}$) are clearly
distinguished. Another feature that is apparent in this
representation is that the magnetization vector passes through
the origin (${\bf M_i}=0$) and reaches the opposite quadrant each
time that (i) a rotation starts moving $\bf n$ away from $\bf H$;
or (ii) the $\bf n \parallel \bf H$ condition is crossed.

Although the CW and CCW curves in Fig. 2(b) are similar (rotated
in $180^{\circ}$ with respect to each other) they also exhibit
some differences. The most obvious one is that the peak at the CD
direction (dotted line) is prominently seen in the CW rotation
(point D), while in the CCW rotation it is partially suppressed
by the flip of ${\bf M_i}$. The flip starts at $\bf H \parallel
\bf n$, and ends at the angle $\Theta_F$ where the fully reversed
critical state is achieved. Making use of the loop analogy, this
requires a field decrease of $\sim 2H^*$, where $H^*(H,T)$ is the
well known full penetration field, then

\begin{equation}
2H^* = H\left( 1-\cos\Theta_F \right)
\label{reverse}
\end{equation}

This analysis indicates that there is a blind range in the
rotation measurements, extending up to an angle $\Theta_F$ from
$\bf n$, where the critical state is not fully developed and thus
$J$ cannot be extracted. Depending on the direction of rotation,
this blind range occurs either in the same quadrant of the CD
(case CCW) or in the opposite (case CW). As $\Theta_F$ decreases
with $H$, in CCW rotations the peak due to the CD is totally
hidden at low fields but can be fully measured at high enough $H$.

The values of $\Theta_F$ are easily obtained from Figure 3(a),
where $M_i$ is plotted as a function of $\Theta$ for the same two
sets of data (CW and CCW) of Fig. 2(b). We observe here that the
agreement between the CW and CCW data is excellent, thus they can
complement each other to eliminate the blind region at low
angles. Estimating $\Theta_F \sim 25^{\circ}$ for the CW rotation
and $\Theta_F \sim 30^{\circ}$ for the CCW case, and using eq.
(\ref{reverse}) we obtain $H^* \sim 370 Oe$ and $\sim 540 Oe$
respectively. We can check the consistency of these estimates in
two ways. First, we know that in a thin sample $H^* \sim J t$.
Combining with the critical state relation $J \sim 60 M_i / W$
(valid for a square platelet) we have $H^* \sim 60 M_i t/W \sim
1.8 M_i$. From the figure we have $M_i(\Theta_F) \sim 200 G$ for
the CW and $\sim 340 G$ for the CCW, so we get $H^* \sim 360 Oe$
and $\sim 610 Oe$ respectively, in very good agreement with the
above estimates. On the other hand, we can compare the values of
$H^*$ obtained from eq. (\ref{reverse}) with those directly
measured in hysteresis loops at the appropriate angles. We have
done so for several temperatures and fields, and we have
systematically obtained very good consistency.

Fig. 3(a) confirms that the condition (\ref{eq:fullp}) is
satisfied in this measurement. In fact, the largest slope
$dM_i/d\Theta \sim 1 kG/rad$, that occurs at $\Theta \sim
33^{\circ}$, implies that $t dJ_c/d\Theta \sim 1.8 kG/rad$, which
is indeed smaller than $H\sin(\Theta) \sim 4.4 kG$. This
condition is also fulfilled in all the cases discussed in the
next section.

In order to compare the data measured by sample rotations with
those resulting from traditional loop measurements, in Fig. 3(a)
we also included $M_i$ values at several $\Theta$ obtained in the
latter way at the same $T$ and $H$ (large open diamonds). The
agreement is very good over the full range of angles, except that
the loop values tend to be somewhat smaller. This is a feature
observed for all measured fields, and can be explained by the
fact that a rotation step is a process that takes only a couple of
seconds, while a field increase and stabilization typically
requires more than 1 minute in our magnetometer, during which the
$\bf M_i$ is already relaxing. Indeed, by performing short
relaxation measurements we have verified that the rotation data
approaches the loop data after 1-2 minutes. This results in
another advantage of the rotations over the loops: measurements
are made closer to the true initial critical state.

Finally, the coincidence of the CW and CCW rotations and the
loops, particularly in the region of the peak due to the CD,
rules out the possibility that vortices lag significantly behind
the direction of $\bf H$ in our rotating sample experiments.
In summary, the information obtained from our rotation measurements
is essentially the same as that provided by hysteresis loops, with several
advantages including the possibility to acquire significantly more data
points for each field. This feature permits a more detailed analysis of
the peak associated with the uniaxial pinning of the CD, as will be shown
in the next section.

\section{Determination of the lock-in angle}

A complete set of rotations at $70K$ for different applied fields
is shown in Figure 3(b). At high fields (above $\sim 6kOe$) a
well-defined peak at the CD direction is observed. At lower
fields ($1kOe \le H \le 5kOe$) the peak progressively broadens and
transforms into a plateau (a certain angular range where
$M_i(\Theta) \sim const.$), while it shifts towards the c-axis. We
had previously reported\cite{silhanek99a} all these features in
YBa$_2$Cu$_3$O$_7$ crystals.

The plateau represents the angular range of applied field over
which it is energetically convenient for the vortices to remain
locked into the columnar defects, thus its angular width is twice
the lock-in angle $\varphi_L$. Below this angle, the vortices are
subject to an invariant (and maximum) pinning force. According to
theoretical models\cite{nel-vin,blatter94}

\begin{equation}
\varphi_L \simeq \frac{ 4\pi\sqrt{2\varepsilon_l\varepsilon_r(T)} }{\Phi_0 H}
\label{eq:lockin}
\end{equation}
where $\varepsilon_l$ is the vortex line tension and $\varepsilon_r(T)$ is the effective pinning energy.

Equation (\ref{eq:lockin}) predicts that $\varphi_L$ should be inversely
proportional to $H$. The improved resolution of the rotation
measurements, that permits a much better determination of the
width of the plateau, allows us to test this dependence. To that
end, we have measured several other sets of data similar to
Figure 3(b), for a wide range of temperatures ($35K$ to $85K$). A
few examples of the observed plateaus are shown in Figure 4.

We then extracted the plateau width for every measurement which
displayed such a feature. This procedure
was done very carefully, including an over-zealous estimate of the
errors involved. The results for all measurable
$\varphi_L$ are plotted as a function of $H^{-1}$
in Figure 5. This figure clearly demonstrates the $H^{-1}$
dependence of $\varphi_L$, as evidenced by the solid lines which are the best linear fits to the data points for each temperature.

According to eq. (\ref{eq:lockin}), the data in fig. 5 should extrapolate to the origin, what is clearly not the case. For all the temperatures where reliable extrapolations can be made ($35K$ to $80K$) the linear fits systematically give a {\it positive} value of $\varphi_L \sim 1.5^{\circ}$ to $3^{\circ}$ at $H^{-1}=0$, which is above the experimental error. There are at least two reasons for this discrepancy. In the first place, we experimentally determine $\varphi_L$ from the intersection of straight lines extrapolated from the plateau and the slopes at both sides of it (see fig. 4). Due to the rounded ends of the plateau, this definition tends to {\it overestimate} $\varphi_L$. Second, the natural splay of the tracks will tend to wash away the expected cusp-like behavior at high fields, thus also contributing to the overestimate of $\varphi_L$. It is clear, on the other hand, that the influence of the splay is not too dramatic, as we indeed observe a rather sharp peak at high fields as seen in fig. 3(b). TRIM calculations show that\cite{suppression}, in our irradiation conditions, the median radian angle of splay slowly increases from zero at the entry surface of the crystal to $\sim 3^{\circ}$ at a depth of $8 \mu m$, and then grows faster to $\sim 6^{\circ}$ at the exit surface.

We now want to analyze whether eq. (\ref{eq:lockin}) provides a satisfactory description of the temperature dependence of the lock-in effect. As this expression does not account for the nonzero extrapolation of $\varphi_L$ discussed in the previous paragraph, it would be incorrect to force a fit through the origin to determine the prefactor of $H^{-1}$. Instead, it is appropriate to identify such prefactor with the slopes $\alpha(T)=d\varphi_L/d(H^{-1})$ of the linear fits. Indeed, the splay of the CD is a geometrical feature independent of H and hence it should only add a constant width to the plateau, without changing its field dependence. To a first approximation, the rounded edges of the plateau will also introduce an additive constant, without significantly affecting the slope. Figure 6 shows the temperature dependence of $\alpha(T)$ (solid symbols). As expected, $\alpha(T)$ decreases with increasing $T$, reflecting the fact that the lock-in angle at fixed $H$ decreases with $T$ due to the reduction of both the line tension and the pinning energy. For a quantitative analysis it is necessary to know the expressions for $\varepsilon_l$ and $\varepsilon_r(T)$. In our experiments the appropriate line tension is that corresponding to in-plane deformations (see pages 1163-1164 in Ref.\cite{blatter94}), $\varepsilon_l=\left(\varepsilon^2\varepsilon_0/\varepsilon(\Theta)\right)\ln\kappa$, where $\varepsilon_0=\left(\Phi_0/4\pi\lambda\right)^2$, the penetration depth $\lambda$ corresponds to ${\bf H}\parallel$c, the mass anisotropy $\varepsilon \ll 1$ and $\varepsilon^2(\Theta)=\cos^2(\Theta)+\varepsilon^2\sin^2(\Theta)$. The temperature dependence of the superconducting parameters appears in $\varepsilon_l$ through $\lambda(T)$. On the other hand, $\varepsilon_r(T)$ is given by\cite{blatter94,nel-vin}

\begin{equation}
\varepsilon_r(T) = \eta \frac{\varepsilon_0}{2} 
\ln\left(1+\frac{r^2}{2\xi^2}\right)\times f(x)
\label{eq:pinenergy}
\end{equation}
where $r \approx 50\AA$ is the radius of the tracks, $\xi$ is the superconducting coherence length, and the dimensionless {\it efficiency factor} $\eta \leq 1$ accounts for the experimental fact that the pinning produced by the CD is smaller than the ideal.\cite{accomodation} Besides the intrinsic temperature dependence of the superconducting parameters, this expression contains an additional temperature dependent factor $f(x)$, known as the {\it entropic smearing} function,  which accounts for the thermal fluctuations of the flux lines. Here $x=T/T_{dp}$, where $T_{dp}$ is a characteristic field-independent {\it depinning temperature}. Combining all these elements, at the track's direction $\Theta=\Theta_{CD}$ we obtain

\begin{equation}
\alpha(T) \approx \frac{\Phi_0\varepsilon}{8\pi\lambda^2} 
\ln\left(1+\frac{r^2}{2\xi^2}\right)\times
\left[\eta  \frac{2\ln\kappa}{\varepsilon(\Theta_D)} f(x) \right]^{1/2}, 
\label{eq:slope2}
\end{equation}

In the original work of Nelson and Vinokur\cite{nel-vin}, where only a {\it short range} pinning potential was considered, the entropic function for $x>1$ was approximately given by $f_{sr}(x) \sim x^2\exp(-2x^2)$. However, according to a further refinement of the model\cite{blatter94}, where the {\it long range} nature of the pinning potential was taken into account, 
this function (for $x>1$) takes the form $f_{lr}(x) \sim \exp(-x)$. 

We can now fit the experimentally determined $\alpha(T)$ using eq. (\ref{eq:slope2}). To that end we use the long range result $f_{lr}(x)$ and fix the reasonably well known superconducting parameters of the material $\varepsilon \approx 1/5$; $\ln\kappa \approx 4$ and $\xi=15\AA /\sqrt{1-t}$ (where $t=T/T_c$). We also assume the usual two-fluid temperature dependence $\lambda(T)=\lambda_L/2\sqrt{1-t^4}$, where $\lambda_L$ is the zero-temperature London penetration depth. The free parameters are then $T_{dp}$ and the combination $\lambda_L/\eta^{1/4}$. The best fit, shown in figure~6 as a solid line, yields $\lambda_L/\eta^{1/4}=360 \AA$ and $T_{dp}=30K$.

Based on the results of figs. 5 and 6, there are a number of considerations that can be made at this point. The first one is that the Bose-glass scenario contained in eqs. (\ref{eq:lockin}) to (\ref{eq:slope2}) provides a quite satisfactory description of the lock-in effect over the whole range of temperature and field of our study. In addition, the obtained $T_{dp}$ is smaller, but still reasonably similar to the value $\sim 41K$ that we had previously found for several YBCO crystals using a completely different experimental method.\cite{accomodation,gaby} This low $T_{dp}$ (well below the initial theoretical expectations) indicates that the efficiency factor $\eta$ is rather small, what is also consistent with the less than optimum $J_c$ observed here and in several previous studies. For low matching fields as that used in the present work, it was estimated\cite{accomodation} that $\eta \sim 0.2 - 0.25$. 

The exact value of $\eta$ has little influence in our estimate of $\lambda_L$, as it only appears as $\eta^{1/4}$. For $\eta=0.2$ and $\eta=1$ we get $\lambda_L=250 \AA$ and $360 \AA$ respectively, a factor of 4 to 5 smaller than the accepted value $\lambda_L \sim 1400\AA$. Zhukov et al.\cite{zhukov97b} had reported a similar discrepancy when studying the lock-in effect by both CD and twin boundaries in YBCO. In a previous study in
YBCO crystals with CD, we had also found that the misalignment between ${\bf B}$ and ${\bf H}$ at low fields (due to anisotropy effects) was well described using a $\lambda_L$ significantly
smaller than the accepted value\cite{silhanek99a}. Thus, this numerical discrepancy appears to be a common result associated to the study of angular dependencies in YBCO-type superconductors
with correlated disorder.

Finally, it is relevant to note that eq. (\ref{eq:lockin}) was derived for the {\it single vortex pinning} regime, which occurs below a temperature dependent accommodation field\cite{blatter94,accomodation} $B^*(T) < B_{\Phi}$, while a large fraction of the data shown in fig. 5 lies above this line, in the {\it collective pinning} regime. Unfortunately, to our knowledge there is no available expression for $\varphi_L \left(H,T \right)$ in the collective regime. Blatter et al.\cite{blatter94} only argued that collective effects should result in a reduction of the lock-in angle. The experimental fact is that eq. (\ref{eq:lockin}) satisfactorily describes both the temperature and field dependence of $\varphi_L$. This suggests that, at least to a first approximation, collective effects in the range of our measurements simply result in a different prefactor in eq. (\ref{eq:lockin}). Clearly, lock-in effects in the collective regime deserve further theoretical study.

\section{Conclusions}

We have measured the irreversible magnetization (${\bf M_i}$) of
an ErBa$_{2}$Cu$_{3}$O$_{7-\delta}$ single crystal with columnar
defects (CD), using an alternative technique based on sample
rotation under a fixed magnetic field. The resulting ${\bf M_i}
(\Theta)$ curves for several temperatures agreed very well
with independent hysteresis loop experiments, showing a peak in
the CD direction at higher fields, while a very well defined
plateau due to the lock-in of the vortices into the CD was
observed at lower fields. The lock-in angle satisfactorily follows
the field and temperature dependence predicted by the Bose-glass scenario.

\section{acknowledgments}

Work partially supported by FAPESP, Brazil, Procs. \#96/01052-7
and \#96/05800-8; ANPCyT, Argentina, PICT 97 No.01120; and CONICET
PIP 4207.

\bibliographystyle{prsty}

\pagebreak

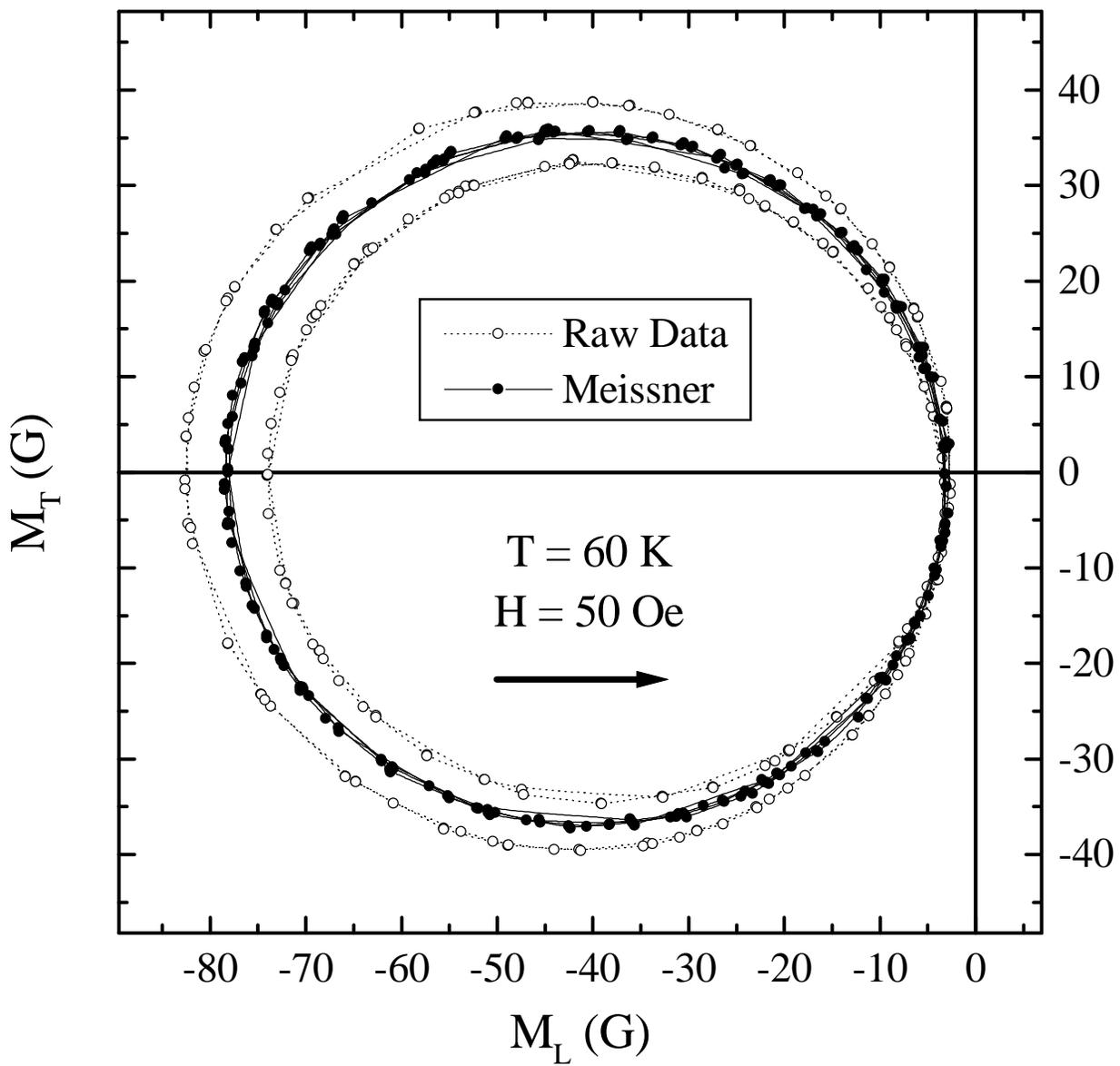
Figure 1. $M_T$ versus $M_L$ polar graph in the Meissner phase for $T=60 K$ and $H=50 Oe$. Open symbols: raw data. Solid symbols: pure Meissner response (the remnant contribution was removed).

\vspace{0.5cm}

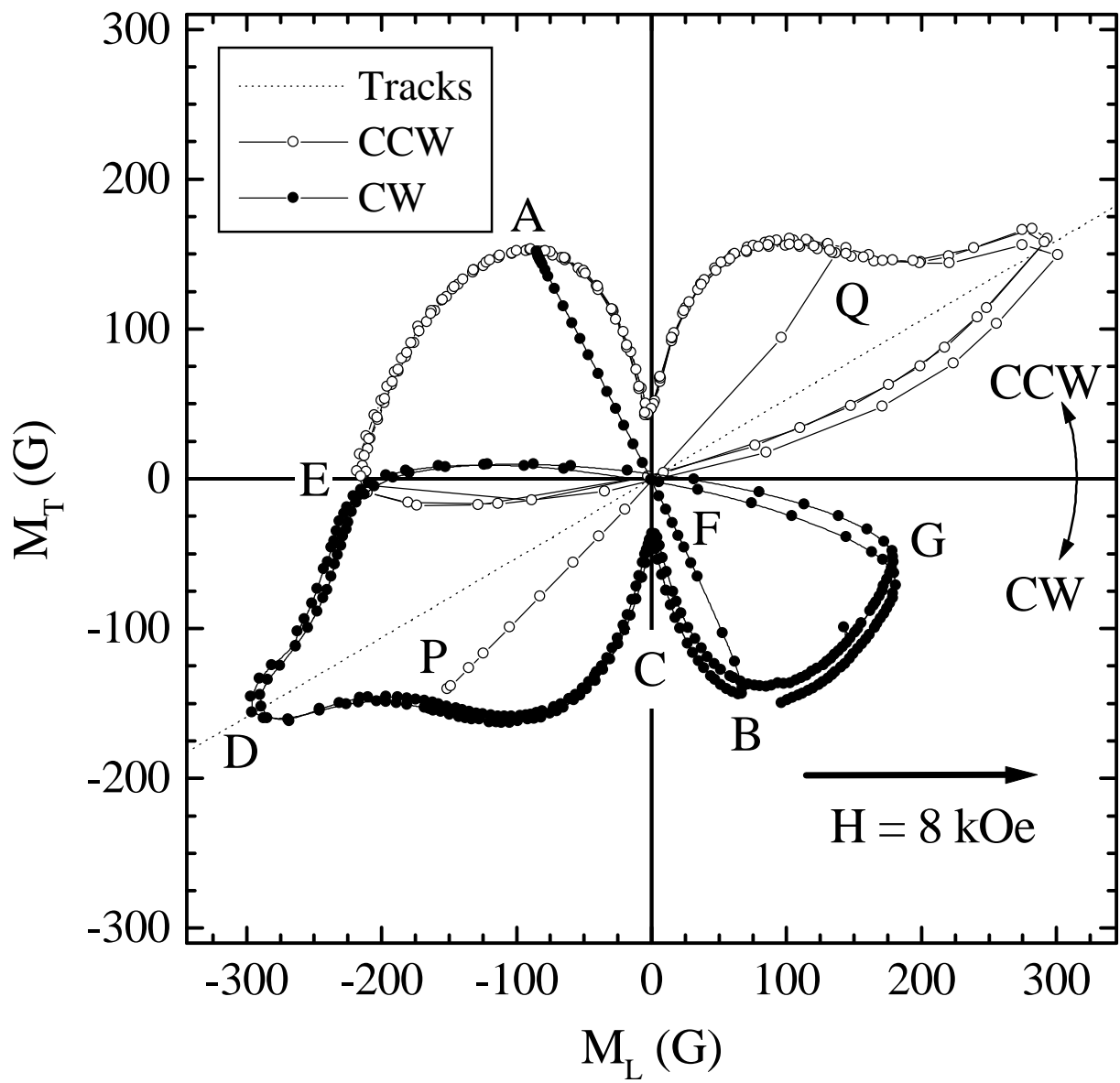
Figure 2. (a) Components of the magnetization vector, $M_T$ and $M_L$, as a function of the angle for $T=70 K$ and $H=8 kOe$. The direction of rotation is such that the conditions
$\bf H \perp c$; $\bf H \parallel$ CD and $\bf H \parallel$ c proceed in that order. (b) $M_T$ vs $M_L$ polar graph of the same CW rotation of (a) (full symbols) together with the CCW rotation.

\vspace{0.5cm}

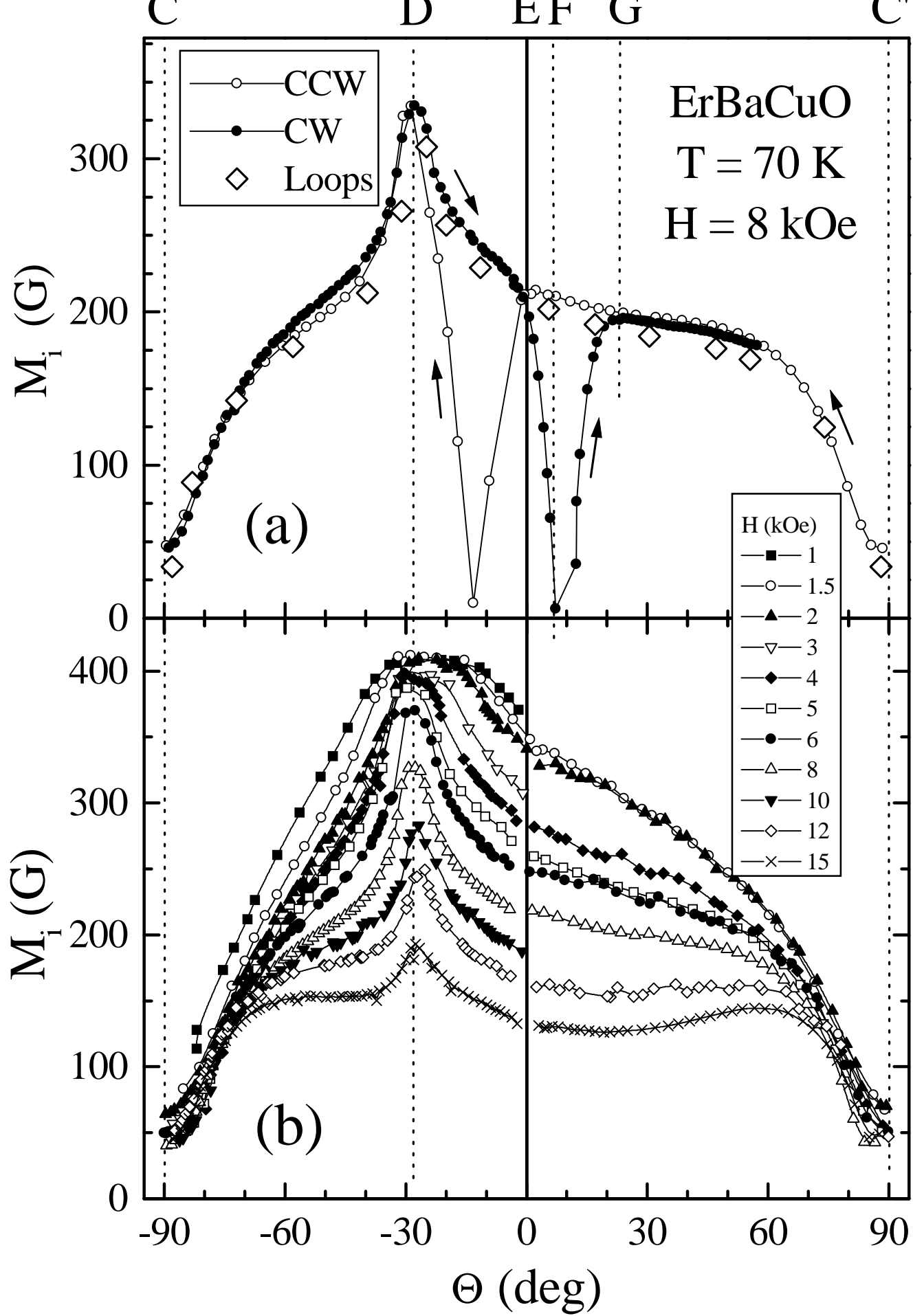
Figure 3. Irreversible magnetization $M_i$ as a function of
$\Theta$ at $T=70K$ for (a) $H=8 kOe$ together with data obtained from hysteresis loop measurements (b) several fields.

\vspace{0.5cm}
Figure 4. Irreversible magnetization $M_i$ as a function of $\Theta$ in the region of the plateau at $T=50 K$ and $70 K$ for several fields.

\vspace{0.5cm}

Figure 5. Lock-in angle $\varphi_{L}$ versus $1/H$ for several temperatures. The straight lines are fits according to equation (\ref{eq:lockin}).

\vspace{0.5cm}
Figure 6. Temperature dependence of the slopes $\alpha(T)=d\varphi_L/d(H^{-1})$ of the linear fits of fig. 5 (full symbols). The solid line is a fit to eq. (\ref{eq:slope2}).

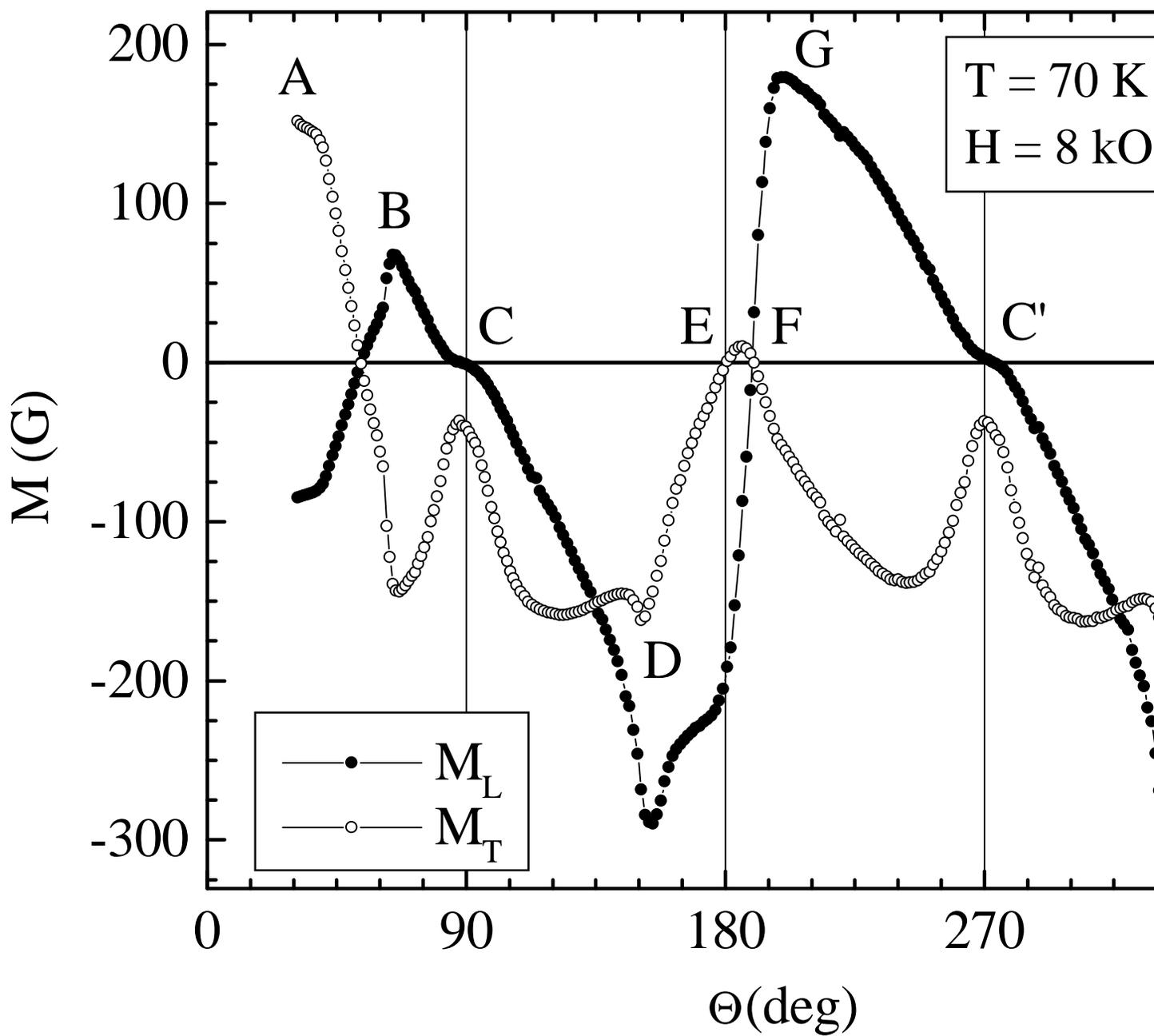
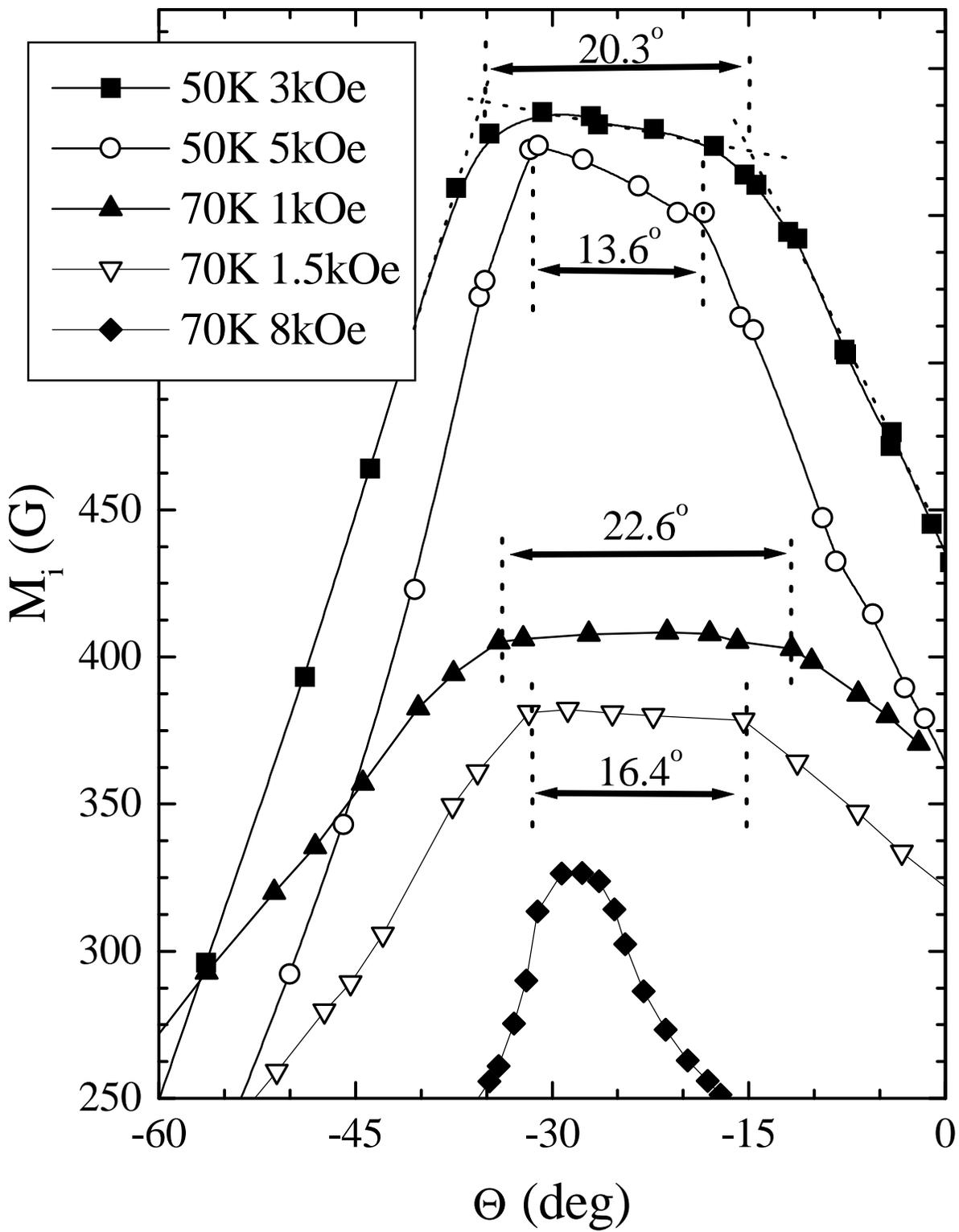
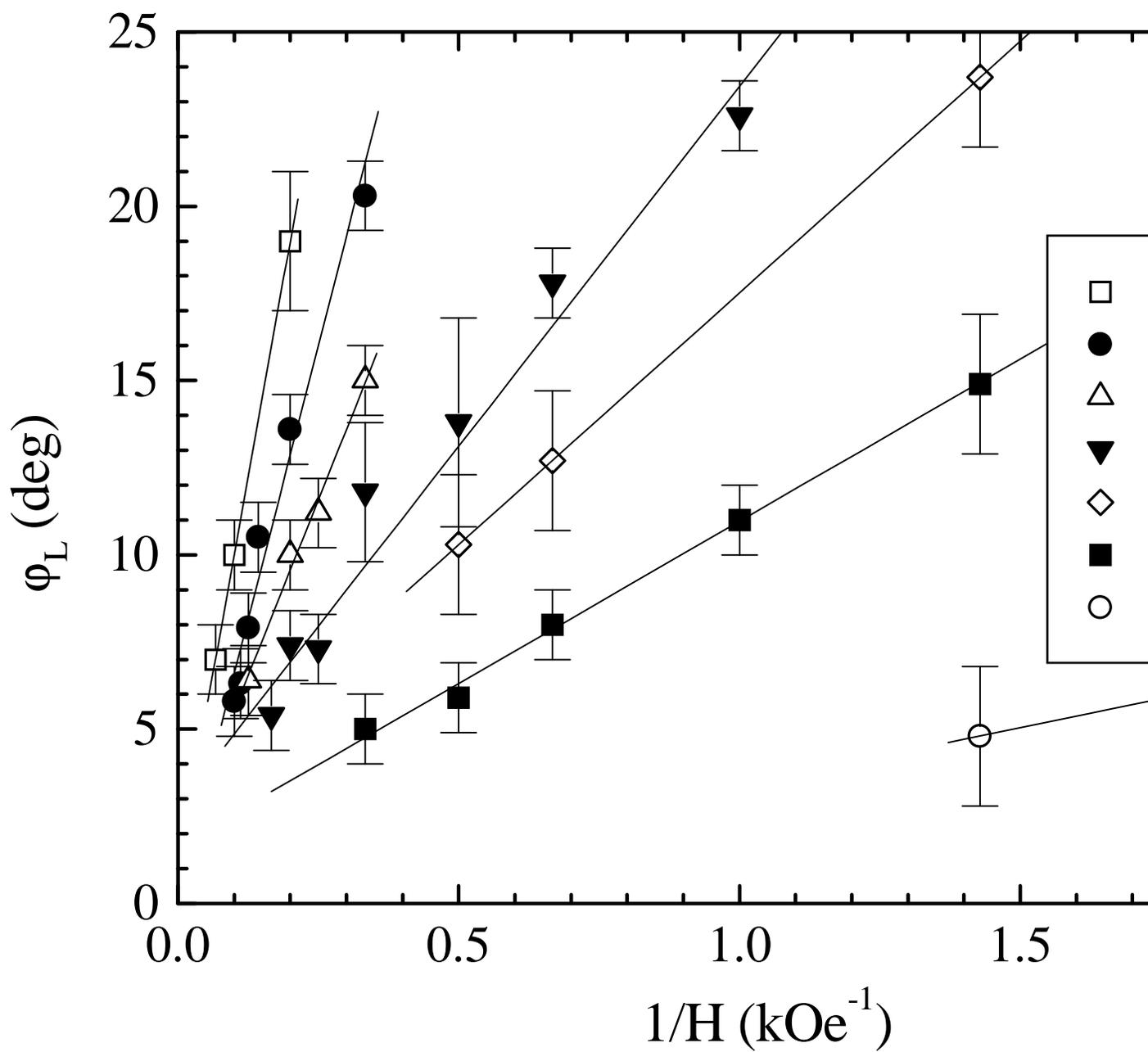
\end{document}